\documentclass[aps,prl,twocolumn,epsfig,amsmath,amssymb]{revtex4}
\usepackage[english]{babel} \usepackage{latexsym}
\usepackage{graphicx} \usepackage{subfigure} \usepackage{epsfig}
\usepackage{psfrag}

\begin{document}

\title{Kelvon-roton instability of vortex lines 
in dipolar Bose-Einstein condensates}
\author{M. Klawunn$^1$, R. Nath$^1$, P. Pedri$^2$ and L. Santos$^1$} 
\affiliation{
\mbox{$^1$Institut f\"ur Theoretische Physik , Leibniz Universit\"at
Hannover, Appelstr. 2, D-30167, Hannover, Germany}\\
\mbox{$^2$Laboratoire de Physique Th\'eorique et Mod\`eles Statistiques,
Universit\'e Paris Sud, 91405 Orsay Cedex, France}\\
}

\begin{abstract}  

The physics of vortex lines in dipolar condensates is studied.
Due to the nonlocality of the dipolar interaction, the 3D 
character of the vortex plays a more important role 
in dipolar gases than in typical short-range interacting ones.
In particular, the dipolar interaction significantly affects 
the stability of the transverse modes of the vortex line. 
Remarkably, in the presence of a periodic potential along the vortex line, 
a roton minimum may develop in the spectrum of transverse modes. We discuss 
the appropriate conditions at which this 
roton minimum  may eventually lead to an instability of the straight 
vortex line, opening new scenarios for vortices in dipolar gases.

\end{abstract}  
 \maketitle



\paragraph{Introduction.-}
When rotated at sufficiently large angular frequency, a superfluid 
develops vortex lines of zero density \cite{Feynman,Landau}, around which  
the circulation is quantized due to the single-valued character of
the corresponding wavefunction \cite{Onsager}. Quantized vortices constitute indeed 
one of the most important consequences 
of superfluidity, playing a fundamental role in various
physical systems, as superconductors \cite{Superconductors} and 
superfluid Helium \cite{Donelly}. Due to the enormous progress in the 
control and manipulation of ultra cold
gases, Bose-Einstein condensates (BECs) offer an extraordinarily controllable
system for the analysis of superfluidity, and in particular quantized vortices 
\cite{Review-Fetter}. Vortices and even vortex lattices have been created in BEC 
in a series of milestone experiments \cite{Matthews99,Madison00,Abo-Shaeer01}.

Vortex lines are indeed 3D structures, which, 
resembling strings, may present transverse excitations, 
which were studied for classical vortices by Lord Kelvin already in 1880, and are 
therefore known as Kelvin modes \cite{Thompson1880}. These excitations 
have been also studied in the context of quantized vortices in superfluids by 
Pitaevskii \cite{Pitaevskii1961}. Interestingly, the dispersion law 
for Kelvin modes at small wave vector $k$ follows a characteristic 
dependence $\epsilon(k) \sim -k^2\ln k\xi$, where $\xi$ is the healing
length. Kelvin modes play an important 
role in the physics of superfluid Helium \cite{Donelly,Ashton79}, 
and even of neutron stars  \cite{Epstein92}. Recently, Kelvin modes were 
experimentally observed in BEC \cite{Bretin2002}.

In spite of being extremely dilute, the properties of ultra cold gases, 
as e.g. superfluidity, are fundamentally determined by the interatomic interactions. 
Up to very recently, typical experiments on ultra cold gases involved 
particles interacting dominantly via a short-range isotropic  
potential, which, due to the very low energies involved, is 
fully determined by the corresponding $s$-wave scattering length. However, recent 
experiments on cold molecules \cite{Molecules}, Rydberg atoms \cite{Rydberg}, and 
atoms with large magnetic moment \cite{Chromium}, open a fascinating new 
research area, namely that of dipolar gases, for which the dipole-dipole 
interaction (DDI) plays a significant or even dominant role. The DDI 
is long-range and anisotropic (partially attractive), 
and leads to fundamentally new physics in condensates 
\cite{Stability,Excitations,Roton}, degenerated Fermi gases \cite{Fermions}, 
and strongly-correlated atomic systems 
\cite{DipLat-FQHE}. It leads to the Einstein-de Haas effect in spinor condensates \cite{EdH}, 
and may be employed for quantum computation \cite{QInf}, and ultra cold chemistry \cite{Chemistry}. 
Time-of-flight experiments in Chromium have allowed for the first observation ever  
of dipolar effects in quantum gases  \cite{Expansion}. Remarkably, 
Feshbach resonances have been very recently employed to reduce 
the scattering length, and as a consequence the role 
of the dipolar interactions has been greatly enhanced \cite{Pfau_new}.

Recently, the physics of rotating dipolar gases has attracted a growing
interest. It has been shown that the critical angular frequency 
for vortex creation may be significantly affected by the DDI
\cite{ODell2005}. In addition, dipolar gases under fast rotation develop
vortex lattices, which due to the DDI may be severely distorted \cite{Pu}, and
even may change its configuration from the usual triangular Abrikosov lattice
into other arrangements \cite{Cooper}. However, to the best of our
knowledge, the previous analysis of vortices in dipolar BEC are constrained to
situations where the 3D character of the vortex is
unimportant.

In this Letter, we analyze the role that the
long-range character of the DDI plays in the physics of vortex lines in
dipolar BEC. Due to this long-range character, different parts of the vortex
line interact via DDI, and hence the 3D character of the vortices
plays a much more important role in dipolar gases than in usual
short-range interacting ones. This has two main consequences. 
On one hand, for a fixed density the vortex core depends on the dimensionality
of the problem, and can hence have in trapped gases 
a non-trivial dependence on the trap geometry. Additionally,  
depending on the dipole orientation, the DDI may significantly enhance or
reduce the stiffness of the line against transverse excitations. Even more
interestingly, under appropriate conditions, discussed in this Letter, the DDI
may induce a minimum in the dispersion law for the transverse modes, i.e. a
Kelvon-roton spectrum. For sufficiently large DDI this minimum may 
reach zero energy, and hence the DDI may destabilize the straight-vortex
configuration, opening the possibility for achieving other ground-state 
vortex configurations.


\paragraph{Effective model.-}
In the following, we consider a dipolar BEC of particles with mass $m$ and 
electric dipole $d$ (the results are equally valid for magnetic dipoles) oriented in the 
$z$-direction by a sufficiently large external field, and that hence 
interact via a dipole-dipole potential:
$V_d(\vec r)=\alpha d^2(1-3\cos^2\theta)/r^3$,
where $\theta$ is the angle formed by the vector 
joining the interacting particles and the dipole direction. The coefficient 
$\alpha$ can be tuned within the range $-1/2\leq\alpha\leq 1$ by rotating the external 
field that orients the dipoles much faster than any other
relevant time scale in the system~\cite{Tuning}. 
At sufficiently low temperatures (and away from shape resonances \cite{Wang2006}) 
the physics of the dipolar BEC is provided by a non-local 
non-linear Schr\"odinger equation (NLSE) of the form:
\begin{eqnarray}
i\hbar \frac{\partial\Psi(\vec{r})}{\partial t}&=&
\left \{ -\frac{\hbar^2\nabla^2}{2m}  +  V_{\rm ol}(z)+ 
g|\Psi(\vec{r})|^2\right\delimiter 0 \nonumber \\ 
&& \left\delimiter 0 \int d \vec{r'} |\Psi(\vec{r'})|^2 V_d(\vec{r}-\vec{r}') \right \}\Psi(\vec{r}),
\label{GPgeneral}
\end{eqnarray}
where $g=4\pi\hbar^2aN/M$, with $a$ the $s$-wave scattering length (we
consider $a>0$). We assume the dipolar BEC in a 1D 
optical lattice formed by counter-propagating lasers, 
$V_{\rm ol}(z)=sE_{R}\sin^2(qz)$, where 
$E_R=\hbar^2q^2/2m$ is the recoil energy and $q$ is the laser wave vector.

In the tight-binding regime (sufficiently strong lattice) 
we can write $\Psi(\vec{r},t)=\sum_jf(z-bj)\psi_j(\vec{\rho},t)$, 
where $b=\pi/q$, $\vec{\rho}=\{x,y\}$ and $f(z)$ is the Wannier function associated to the 
lowest energy band. Substituting this Ansatz in Eq.~(\ref{GPgeneral}) we
obtain a discrete non-linear Schr\"odinger equation. We may then 
return to a continuous equation, where the lattice is taken
into account in an effective mass along the lattice direction and in the renormalization
of the coupling constant \cite{Meret,nota}: 
\begin{eqnarray}
i\hbar \frac{\partial\Psi(\vec{r})}{\partial t}&=&
\left \{ -\frac{\hbar^2\nabla_\perp^2}{2m}- \frac{\hbar^2\nabla_z^2}{2m^*} + \tilde{g}|\Psi(\vec{r})|^2+\right\delimiter 0 \nonumber \\ 
&& \left\delimiter 0 \int d \vec{r'} |\Psi(\vec{r'})|^2 V_d(\vec{r}-\vec{r}') \right \}\Psi(\vec{r}),
\label{GPeffe}
\end{eqnarray}
where $\tilde{g}=bg\int f(z)^4dz+g_d\mathcal{C}$ \cite{nota}, 
with $g_d=\alpha 8\pi d^2/3$, 
$m^*=\hbar^2/2b^2J$ is the effective mass, and 
$J=\int f(z)[-(\hbar^2/2m)\partial^2_z+V_{\rm ol}(z)]f(z+b)dz$. 
Note that the discreteness of the lattice is unimportant, i.e. 
Eq. (\ref{GPeffe}) is valid, if $k_z\ll2\pi/b$, being $k_z$ 
the $z$-momentum.
In the following we use the convenient dimensionless parameter
$\beta=g_d/\tilde{g}$, that   
characterizes the strength of the DDI compared
to the short-range interaction. It will be also useful in the following 
the Fourier transform of the DDI: 
$\tilde V_d(\vec k)=g_d[3\cos^2\theta_k-1]/2$, 
with $\cos^2\theta_k=k_z^2/|\vec k|^2$.  

\paragraph{Homogeneous solution.-}
We consider first an homogeneous solution of the form 
$\Psi_0(\vec r, t)=\sqrt{\bar n}\exp [-i\mu t/\hbar]$, where $\bar n$ denotes the 
condensate density, and $\mu=(g+\tilde V_d(0))\bar n$ is the 
chemical potential. From the corresponding Bogoliubov-de Gennes (BdG) 
equations one obtains that the energy $\epsilon(\vec k)$ corresponding 
to an excitation of wave number $\vec k$ fulfills: 
$
\epsilon(\vec k)^2=E_{\rm kin}(\vec k)[E_{\rm kin}(\vec k)+E_{\rm int}(\vec k)]
$
where $E_{\rm kin}(\vec k)=\hbar k_\rho^2/2m+\hbar k_z^2/2m^*$ is the kinetic energy, and 
$E_{int}(\vec k)=2(g+\tilde V_d(\vec k))\bar n$. Stable phonons (i.e. 
excitations at low $k$) are only possible if 
$E_{int}>0$ for all $\vec k$, i.e. if  
$2+\beta (3\cos^2\theta_k-1)>0$. 
If $g_d>0$  phonons with $\vec k$ lying on the $xy$ plane are
unstable if $\beta>2$, while for $g_d<0$ 
phonons with $\vec k$ along $z$ are unstable if $\beta<-1$. 
Hence absolute phonon stability demands $-1<\beta<2$. 

\begin{figure}
\begin{center}
\includegraphics[width=5.0cm]{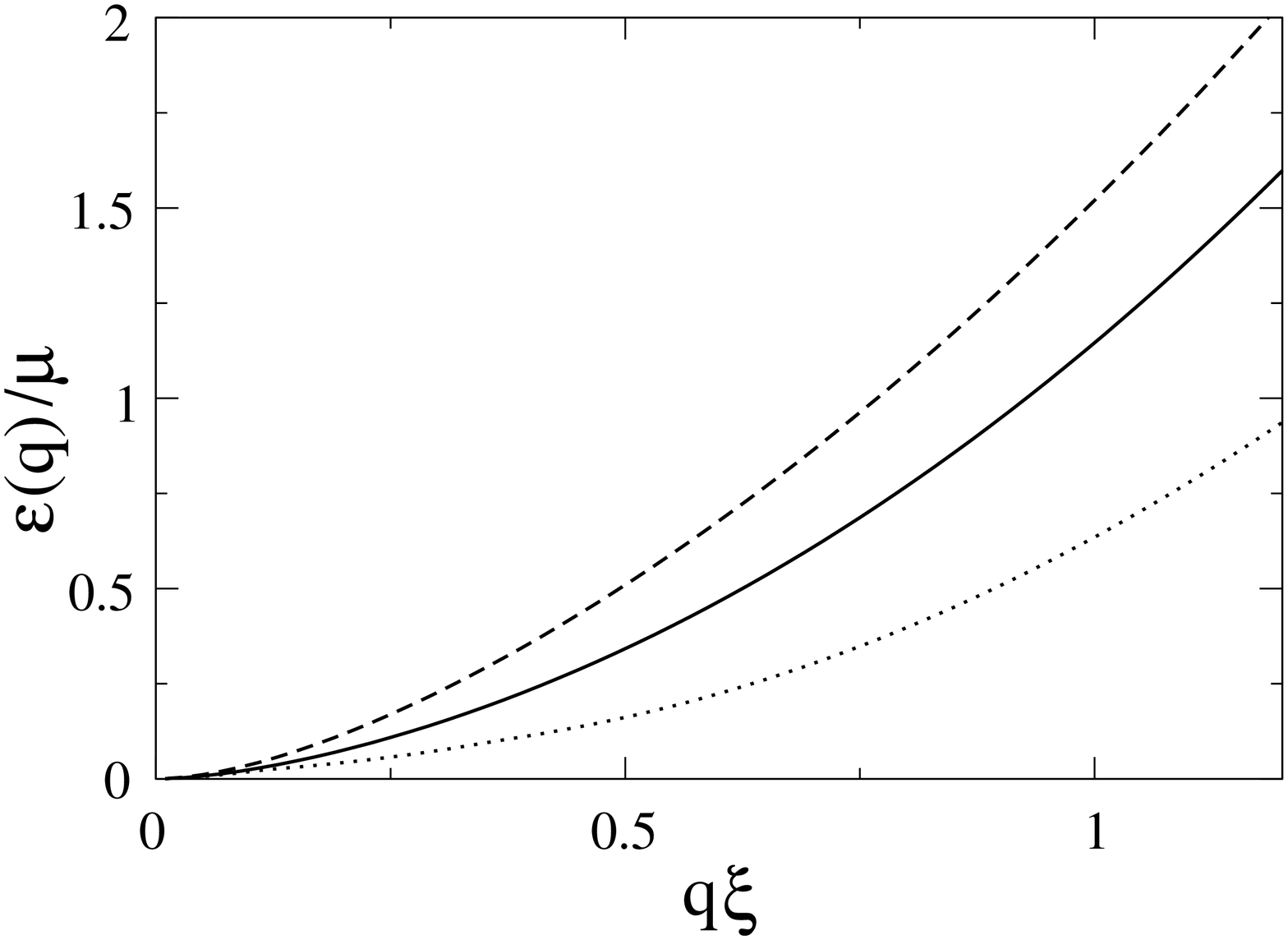}
\end{center} 
\vspace*{-0.5cm}
\caption{Dispersion $\epsilon(q)$ as a function of $q\xi$, in the absence 
of an additional optical lattice, and various
  values of $\beta=0$ (solid), $1.0$ (dashed), and $-0.98$ (dotted).}
\label{fig:1}
\end{figure}
\begin{figure}
\begin{center}
\includegraphics[width=5.0cm]{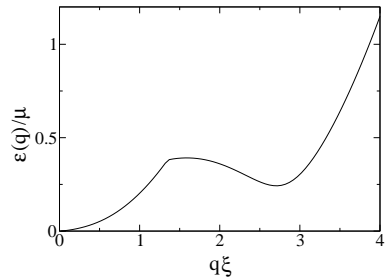}
\end{center} 
\vspace*{-0.5cm}
\caption{Dispersion $\epsilon(q)$ as a function of $q\xi$ for $m/m^*=0.23$, 
and $\beta=-0.59$.}
\vspace*{-0.2cm}
\label{fig:2}
\end{figure}


\paragraph{Vortex solution.-}
We consider at this point a condensate with a straight vortex line along the 
$z$-direction. The corresponding ground-state wavefunction is of the form 
$\Psi_0(\vec r,t)=\phi_0(\rho)\exp(i\varphi)\exp [-i\mu t/\hbar]$, where $\varphi$ is
the polar angle on the $xy$ plane. The function $\phi_0(\rho)$ fulfills 
\begin{equation}
\mu\phi_0(\rho)=\frac{\hbar^2}{2m}\left(-D_\rho+\frac{1}{\rho^2}\right)\phi_0(\rho)+\bar{g}|\phi_0(\rho)|^2\phi_0(\rho),
\end{equation}
where $D_\rho=\frac{1}{\rho}\partial_\rho\rho\partial_\rho$ and 
$\bar{g}=\tilde{g}-g_d/2$.
Note that since $\Psi_0$ is independent of
$z$, the DDI just leads to a regularization of the contact interaction. 
The density of the vortex core is given by $|\phi_0(\rho)|^2$, which goes to
zero at $\rho=0$, and becomes equal to the bulk
density $\bar{n}$ at distances larger than the corresponding healing length 
$\xi=\hbar/\sqrt{m\bar{g} \bar{n}}$. Note that due to the 
regularization of the contact interaction, the size $\xi$ of
the vortex core depends on the DDI. In particular, this dependence is exactly
the opposite as that expected for 2D vortices, since in 2D
similar arguments provide $\bar g=g+g_d$. Hence, even for equal densities, and due
to the long-range character of the DDI, the cores
of a 2D vortex and a 3D vortex line can be remarkably different in a dipolar
gas, differing significantly from the behavior of short-range interacting
gases, where both 2D and 3D vortices would have the same core size.

The effects of the long-range character of the DDI become even more relevant
in the physics of Kelvin modes. In order to analyze these modes, we
consider a perturbation of the straight vortex solution of the form 
$\Psi(\vec r,t)=\Psi_0(\vec r,t)+\chi(\vec r,t)\exp [i (\varphi-\mu t/\hbar)]$, 
where \cite{Pitaevskii1961} 
$\chi(\vec r,t)=\sum_l [ u_l (\rho) \exp [i(\varphi l+qz-\epsilon t/\hbar)]-v_l(\rho)^*
\exp [-i(\varphi l+qz-\epsilon^* t/\hbar)]]$.
Introducing this Ansatz into~(\ref{GPeffe}) and linearizing in $\chi$, one obtains
the corresponding BdG equations:
\begin{widetext}
\begin{eqnarray}
\epsilon u_l(\rho) &=& \left[ \frac{\hbar^2}{2m}\left(-D_\rho
 + \frac{(l+1)^2}{\rho^2}+\frac{m}{m^*}q^2\right) -\mu+ 2\bar g \psi_0(\rho)^2 \right]u_l(\rho) - \bar g \psi_0(\rho)^2 v_l(\rho) \nonumber \\
&+&\frac{3\beta}{2-\beta}\bar g  \int_0^\infty d\rho' \rho' \psi_0(\rho') \psi_0(\rho)
\left[u_l(\rho')-v_l(\rho') \right]F(k\rho,k\rho') \label{BdG1} \\
\epsilon v_l(\rho) &=& 
-\left[ \frac{\hbar^2}{2m}\left(-D_\rho
 + \frac{(l-1)^2}{\rho^2}+\frac{m}{m^*}q^2\right) -\mu+ 2\bar g \psi_0(\rho)^2 \right]v_l(\rho)
+ \bar g \psi_0(\rho)^2 u_l(\rho) \nonumber \\
&+&\frac{3\beta}{2-\beta}\bar g  \int_0^\infty d\rho' \rho' \psi_0(\rho') \psi_0(\rho)
\left[u_l(\rho')-v_l(\rho') \right]F_l(k\rho,k\rho'), \label{BdG2}
\end{eqnarray}
\end{widetext}
with $F_l(x,x')=I_l(x_<)K_l(x_>)$ ,
where $I_l$ and $K_l$ are modified Bessel functions, and 
$x_>=$max$(x,x')$, $x_<=$min$(x,x')$. For every $q$ we determine the lowest 
eigenenergy $\epsilon$, that provides the dispersion law discussed below.
The first line at the rhs of Eqs. (\ref{BdG1}) and (\ref{BdG2}) is exactly the
same as that expected for a vortex in a short-range interacting BEC
\cite{Pitaevskii1961}, but with
the regularized value $\bar g$. 
The last term at the rhs of both equations
is directly linked with the long-range character of the DDI and, as we show
below, leads to novel phenomena in the physics of Kelvin modes ($l=1$) in dipolar
BECs. 

In absence of DDI (or equivalently from Eqs.~(\ref{BdG1})
and (\ref{BdG2}) without the last integral term) the dispersion law
at low momenta ($q\xi\ll 1$) is provided by the well-known expression
$\epsilon(q)=-(\hbar^2q^2/2m^*)\ln \left[ (m/m^*)^{1/2}q\xi\right]$.
The integral terms of Eqs. (\ref{BdG1}) and (\ref{BdG2}) significantly
modify the Kelvon spectrum in a different way depending whether $\beta$ is 
positive or negative. In order to isolate the effect of the DDI on
the core size with respect to the effect of the integral terms in Eqs. 
(\ref{BdG1}) and (\ref{BdG2}) we fix $\bar g$ and change
the parameter $\beta$ which is proportional to the dipole-dipole coupling constant. 
Fig.~\ref{fig:1} shows that for increasing $\beta>0$ the
excitation energy clearly increases, i.e. the vortex line becomes 
stiffer against transverse modulations. 

In order to obtain an intuitive picture of why this is so, one may sketch the 
vortex core as a 1D chain of dipolar holes. Dipolar holes interact in exactly
the same way as dipolar particles, and hence maximally attract each
other when aligned along the dipole direction, i.e. the $z$-direction. In this 
sense, the configuration of minimal dipolar energy is precisely that of a
straight vortex line along $z$. A wiggling of the line produces a displacement
of the dipolar holes to the side, and hence an increase of the dipolar
energy. As a consequence, the DDI leads to an enhanced stiffness of the 
vortex line.
From this intuitively transparent picture, we can easily understand that
exactly the opposite occurs when $\beta<0$. In that case, the dipolar holes 
maximally repel each other when aligned along the $z$-direction. Hence 
it is expected that the dipolar energy is minimized when departing
from the straight vortex configuration. This results in a reduced energy of
the excitations for $\beta<0$ (
Fig.~\ref{fig:1}), i.e. it becomes easier to wiggle the vortex line. 

\begin{figure}
\begin{center}
\includegraphics[width=5.5cm]{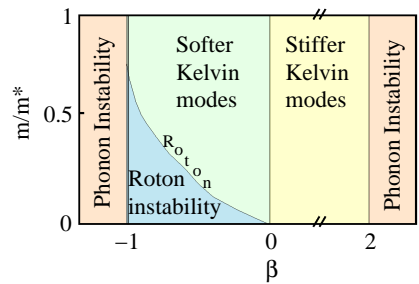}
\end{center} 
\vspace*{-0.6cm}
\caption{Stable/unstable regimes for straight vortex lines.}
\vspace*{-0.3cm}
\label{fig:3}
\end{figure}

In principle, a sufficiently large DDI could even destabilize the straight 
vortex line. However, in the absence of an additional optical lattice 
along the $z$-direction, the destabilization would occur for values of $\beta<-1$, 
i.e. in the regime of phonon instability in 
which the whole dipolar BEC is unstable against local collapses. 
An increase of the potential depth of the additional lattice leads to a reduction of  
the role of the kinetic energy term $mq^2/m^*$ in Eqs. (\ref{BdG1}) and (\ref{BdG2}) that 
enhances the effect of the dipolar interaction on the
dispersion law. As a consequence, 
as shown in Fig.~\ref{fig:2}, in addition to the $-k^2\ln (k\xi)$ dependence
at low momenta, a minimum eventually appears at intermediate momenta, i.e. the spectrum
develops a Kelvon-roton character. Note that, typically, 
the dispersion law presents a relatively 
abrupt change in its curvature for a given value 
of $q$ ($\simeq 1.4$ in Fig.~\ref{fig:2}). 
Interestingly this feature is given by an avoided level-crossing of the 
lowest eigenvalues $\epsilon(q)$. As a consequence the character of the 
lowest Bogoliubov modes changes at this value of $q$.
For a sufficiently small $(m/m^*)_{cr}$ the roton minimum 
eventually reaches zero energy, and the straight-vortex line becomes
unstable against a novel type of instability. Fig.~\ref{fig:3} shows  
as a function of $(m/m^*)$ and $\beta$, the different regimes for the straight-vortex 
line.

Roton minima occur in the dispersion law of superfluid Helium \cite{Feynman}, 
and have been also predicted in trapped dipolar gases under a strong 1D and 2D 
confinement \cite{Roton}. In those cases, a phonon-roton
spectrum occurs related to excitations of the homogeneous BEC, and the 
eventual roton instability is linked to the appearance of density
fluctuations (which are expected to lead to local collapses
\cite{Komineas-GoraPrivate}). Although the Kelvon-roton spectrum discussed in this
Letter is also induced by the DDI, it fundamentally differs from the
phonon-roton case. In this case, the roton minimum occurs due to the
long-range interaction between different positions in the vortex line. In
addition, the eventual roton instability indicates that the straight vortex
line ceases to be the ground-state configuration, opening the exciting possibility to
achieve ground-state helicoidal vortex line configurations. The analysis of this possibility
will be the subject of further investigations.


\paragraph{Summary.-} 
We have shown that, due to the nonlocal interaction between different parts
of the vortex line, the 3D character of the vortices is much more crucial 
in dipolar BECs than in short-range interacting ones. 
The size of the vortex core depends significantly on
the system dimensionality, which indicates that, 
although we have restricted to homogeneous gases, 
vortex lines in trapped dipolar BECs may have a non-trivial dependence 
on the trap geometry. 
On the other hand, and depending on the dipole orientation, the DDI 
may significantly increase or decrease the stiffness of the 
line against Kelvin excitations.
Interestingly, under appropriate conditions (additional 
optical lattice along the vortex line), a Kelvon-roton dispersion law occurs. 
For sufficiently large DDI a roton instability of the straight vortex line develops, 
which may preclude the appearance of a new type of helicoidal ground-state 
configuration of the vortex line.

\acknowledgements
Conversations with G. V. Shlyapnikov are acknowledged. 
This work was supported by the DFG (SFB-TR21, SFB407, SPP1116).
the Minist\`ere da la Recherche (grant ACI 10
Nanoscience 201), the ANR (grants NT05-2-42103 and
05-Nano-008-02), and the IFRAF Institute.



\begin{thebibliography}{99}
\bibitem{nota}  
${\mathcal C}\simeq \sum_{j\ne 0} |\tilde f(2\pi j/d)|^2$, where 
$\tilde f$ is the Fourier-transform of $f(z)$.
\end{thebibliography}
\end{document}